# Cryogenics and the use of superfluid helium in high-energy particle accelerators (1980 – 2000)

Ph. Lebrun

CERN Emeritus

**Abstract.** The period 1980 – 2000 saw the impressive development of applied superconductivity in high-energy particle accelerators, from single components to long strings of superconducting magnets and high-frequency acceleration cavities. Large and powerful cryogenic systems were designed ancillary to superconducting devices operating generally close to the normal boiling point of helium, but also above 4.2 K in supercritical and below 2 K in superfluid. Low-temperature operation in accelerators also involves considerations of ultra-high vacuum, limited stored energy and beam stability. We recall the rationale for cryogenics in high-energy particle accelerators and review its development over the period of interest, with reference to the main engineering domains of cryostat design and heat loads, cooling schemes, efficient power refrigeration and cryogenic fluid management. In view of its importance and novelty, a specific section is devoted to the developments that led to the LHC at CERN.

# Cryogenics and the use of superfluid helium in high-energy particle accelerators (1980 – 2000)

Ph. Lebrun

CERN Emeritus

**Abstract**. The period 1980 – 2000 saw the impressive development of applied superconductivity in high-energy particle accelerators, from single components to long strings of superconducting magnets and high-frequency acceleration cavities. Large and powerful cryogenic systems were designed ancillary to superconducting devices operating generally close to the normal boiling point of helium, but also above 4.2 K in supercritical and below 2 K in superfluid. Low-temperature operation in accelerators also involves considerations of ultra-high vacuum, limited stored energy and beam stability. We recall the rationale for cryogenics in high-energy particle accelerators and review its development over the period of interest, with reference to the main engineering domains of cryostat design and heat loads, cooling schemes, efficient power refrigeration and cryogenic fluid management. In view of its importance and novelty, a specific section is devoted to the developments that led to the LHC at CERN.

## 1. Introduction

Cryogenics, together with applied superconductivity, has become an enabling technology of particle accelerators for high-energy physics, contributing to their sustained development since the 1980s. Conversely, this development has produced new challenges and markets for cryogenics, resulting in a fruitful symbiotic relation which materialized in significant technology transfer and technical progress. This is exemplified by the time evolution of the total cryogenic refrigeration capacity at liquid helium temperature installed at CERN, the European Organization for Nuclear Research in Geneva, Switzerland (figure 1). Although some fairly large helium cryogenic plants have been in use at CERN since the 1960s, mainly feeding large bubble chamber and detector magnets, it is only the generalized application of superconductivity in particle accelerators – RF acceleration cavities and high-field bending and focusing magnets – that has led to the expansion of cryogenics, with kilometre-long strings of helium-cooled devices, powerful and efficient refrigerators and superfluid helium used in high tonnage as cooling medium.

It is therefore interesting to analyze this transition, taking stock of the technical progress stimulated by the diverse large accelerator projects of this period and trying to understand the evolutionary processes at work. Before discussing historical development, let us first recall the rationale for the use of cryogenics in high-energy particle accelerators [1].

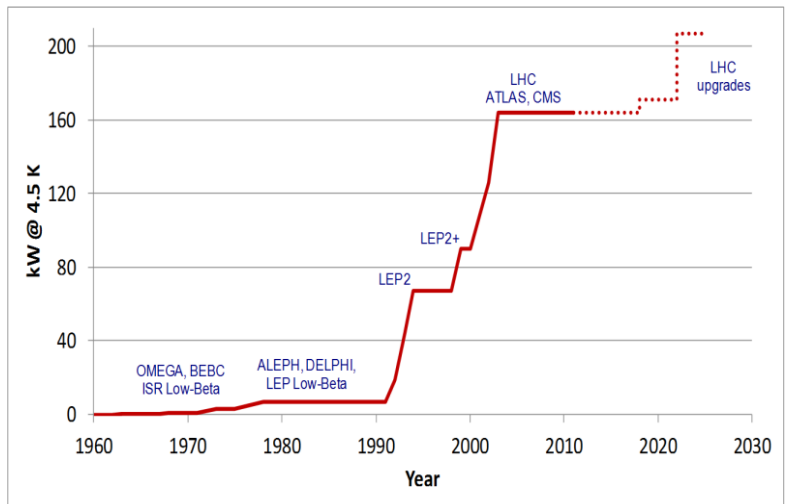

**Figure 1.** Installed cryogenic refrigeration capacity at CERN

## 2. The rationale for cryogenics in particle accelerators

The primary use of cryogenics in particle accelerators is to cool their superconducting components. The absence of electrical d.c. resistance, or the limited a.c. dissipation in superconductors opens the way to produce higher fields and thus reach higher beam energy, while containing the dimensions of particle accelerators. This was historically the first reason to bring superconducting magnets into circular accelerators. More recently, with the advent of large machines, electrical power consumption became a serious issue, and the choice of superconducting technology was also driven by energy efficiency. This applies particularly to "continuous-wave" linear accelerators, where most of the power is dissipated in the walls of the accelerating high-frequency structures, but also to the magnet systems of very large circular machines.

Circular accelerators have developed along two different lines to handle ever stiffer high-energy beams (figure 2): increasing their diameter, and hence the radius of curvature in bending magnets, and increasing the field produced by these magnets. Normal-conducting magnets are iron-dominated, and hence practically limited to below 2 T by the saturation of their magnetic iron yoke. In superconducting magnets, the ampere-turns cost little, and the field is directly produced by the current distribution, without the need to concentrate the flux in an iron yoke. The magnets are therefore limited by the "critical surface" in the temperature/field/current density space of the superconductor used to wind their coils or equivalently, at any given temperature, by the "critical curve" in the field/current density plane. To obtain sufficient current-carrying capacity at high field, the superconductor must then be operated at a fraction of its "critical temperature", in practice half of it or below. Hence normal-boiling helium at 4.2 K appears an adequate coolant for most magnets wound with niobium-titanium superconductor, which has a critical temperature of about 9.5 K. If, however, one wishes to produce fields in the 8 T to 10 T range with niobium-titanium, one must operate the magnets at lower temperature, e.g. below 2 K in superfluid helium in order to maintain sufficient current-carrying capacity at high field. Conversely, normal-boiling nitrogen at 77 K appears not cold enough to cool magnets made of today's high-temperature superconductors. The d.c. superconducting magnet system of a particle collider does not show steady-state dissipation other than the power consumption of the cryogenic refrigeration system, which scales approximately with the circumference of the machine, irrespective of the bending field. The specific power consumption (kW per GeV of beam energy) of the superconducting magnet system – including cryogenic refrigeration – therefore scales as the inverse of the field (figure 3). High-energy circular machines thus need superconductivity both for reasons of compactness and energy efficiency.

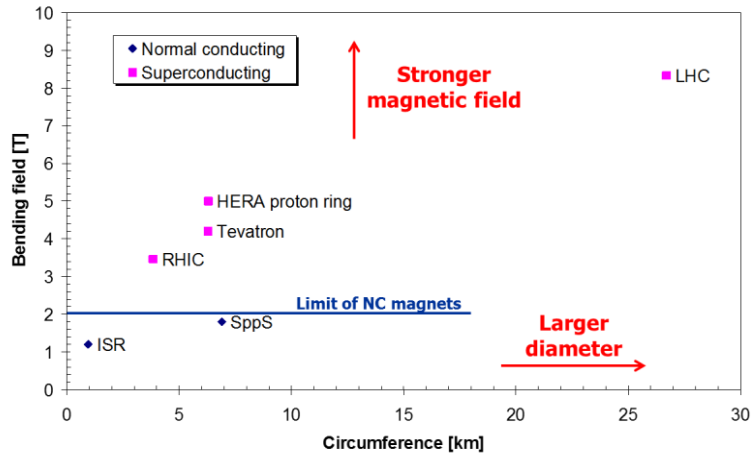

**Figure 2.** Development of circular hadron colliders

The case of linear accelerators, mostly composed of high-frequency accelerating cavities, is different. Such cavities are resonators for which one wishes to maximize the quality factor to reduce the power dissipation in the wall, characterized by its surface resistance. While copper cavities at room temperature show typical quality factors of few $10^4$, superconducting cavities can reach quality factors of several $10^9$, with a reduction in power dissipation more than offsetting the power consumption of the cryogenic refrigerator. This applies particularly for "continuous-wave" high-frequency systems, in which the cavities are continuously energized. Moreover, the surface resistance of superconducting cavities, which controls the quality factor and hence the power dissipation, has a component which scales with the square of frequency and the exponential of the ratio of critical to operating temperature. Consequently, and despite the higher thermodynamic cost of refrigeration at lower temperature, it may be advantageous to operate high-frequency superconducting cavities at temperatures lower than the normal boiling point of helium, e.g. in superfluid helium below 2.2 K. With today's niobium cavities, this typically applies for frequencies above 700 to 800 MHz. The use of niobium with lower residual resistance displaces the optimum towards lower temperatures. Conversely, building the cavities out of higher-temperature superconductors, or coating their wall with such materials, would reduce the exponential term when operating them in normal boiling helium.

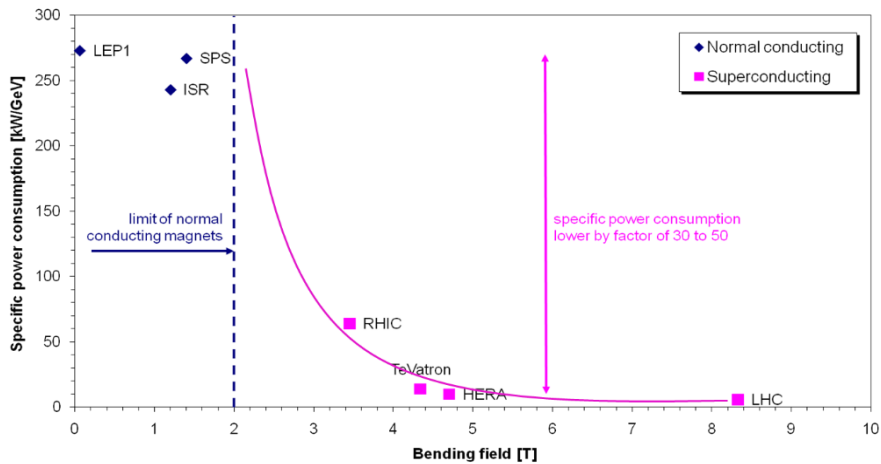

**Figure 3.** Specific power consumption of circular colliders

Other arguments for the use of cryogenics in particle accelerators are not driven by superconductivity, but by the interactions of the circulating beams with their environment, i.e. the wall of the beam pipe and the residual gas pressure in it. The circulation of charged particles induces currents in the metallic walls of the beam pipe, producing fields which act on the particles. This interaction, characterized by an impedance proportional to wall resistivity, leads to power dissipation and in some cases to beam instabilities. This effect is important in large accelerators with small aperture and can be compensated by feedback provided the rise time of the instability is long enough, i.e. the impedance can be kept low. This sets constraints on the choice of material and operating temperature for the first wall seen by the beam. In the Large Hadron Collider (LHC), the beam screen is coated with copper and maintained below 20 K [2].

Maintaining the beam pipe at low temperature also provides excellent cryopumping of most residual gas species. The saturation pressure of all gases, except helium, vanish at cryogenic temperatures (figure 4), so superconducting accelerators which must be cooled at liquid helium temperature may benefit from this feature. In cases when the residual pressure in the beam enclosure must be kept very low, such as accelerators of highly charged ions, a cold beam pipe may become the driving factor for the use of cryogenics.

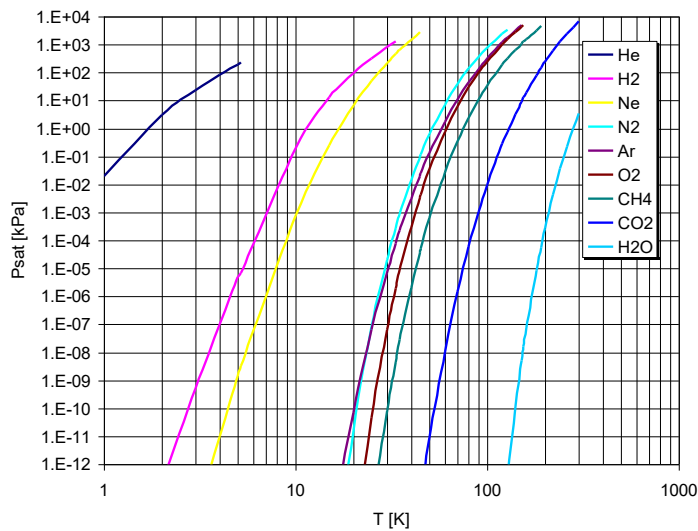

**Figure 4.** Saturation pressure of gases as function of temperature

### 3. Superfluid helium as a technical coolant

While the second liquid phase of helium had been observed soon after its first liquefaction by H. Kamerlingh Onnes in Leiden in 1908 and unsuccessful attempts to solidify it [3], it took thirty years to discover its peculiar transport properties, simultaneously by P. Kapitza in Moscow [4] – who coined the word "superfluidity" by analogy with superconductivity – as well as by J.F. Allen and A.D. Misener in Cambridge [5]. Another half century was necessary to bring superfluid helium from laboratory curiosity and advanced research topic in condensed matter physics, to technical coolant for superconducting devices [6].

The prime reason for cooling with superfluid helium is evidently the lower temperature than normal boiling helium, providing more design space under the critical surface of superconductors, i.e. more current-carrying capacity at high field for d.c. electromagnets or lower power dissipation for a.c. applications such as RF cavities. Additionally, the cryogenic system should be designed to reap full benefit of the exceptional transport properties of the coolant. The low effective viscosity – typically a hundred times lower that that of water at the normal boiling point – allows good permeation within the compound structure of the magnet windings and good wetting of the superconductor. The high specific heat – 2000 times that of the superconductor per unit volume – is an important asset for stabilizing it against thermal disturbances. The high thermal conductivity – typically 1000 times that of OFHC copper between 1.8 K and 2.0 K – can extract heat efficiently from the magnet windings, provided sufficient percolation paths are maintained. It can also ensure quasi-isothermal helium baths and transport heat over limited distances. It is however insufficient to cool kilometre-long strings of magnets by sole conduction [7]. Superfluid helium can be used as coolant either at saturation (i.e. below 5 kPa), or as subcooled liquid at pressure close to atmospheric – "pressurized superfluid helium" – (figure 5). The latter limits the risk of air inleaks and contamination, as well as electrical breakdown due to the bad dielectric characteristics of low-pressure helium vapour (Paschen curve). It also provides better heat buffering thanks to higher enthalpy margin from the working point up to the lambda line. We shall see in the following how superfluid helium cooling is implemented in the cryogenic systems of some particle accelerators.

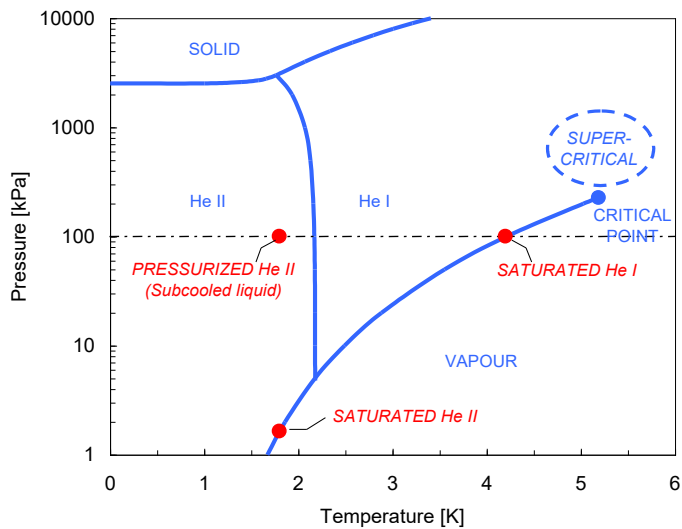

**Figure 5.** Phase diagram of helium, showing saturated and pressurized superfluid domains

## 4. Projects and progress

Although the idea of using superconductors to build high-field magnets had been formulated by H. Kamerlingh Onnes soon after the discovery of superconductivity, it is only with the advent of type-II superconductors that it could be put into practice: a 1.5 T superconducting magnet wound with molybdenum-rhenium alloy was built and patented by J. Kunzler in 1960, and the niobium-titanium alloys which became the workhorse of applied superconductivity were discovered in 1961 [8].

The potential of superconductivity for improving linear accelerators, including the option of superfluid helium cooling of the high-frequency cavities [9] was identified in the 1960s. In the same period, the benefits of using superconducting magnets in proton synchrotrons were clearly formulated and quantified, in relation to future projects of the time [10]. In the following, we present projects which can be considered as important milestones in the development of cryogenics in particle accelerators. Rather than attempting to give a complete description of these projects or of their cryogenic system, we single out their specific features which have resulted in progress in the field. This presentation is evidently not free of personal bias.

The first system of superconducting magnets to be routinely operated in an accelerator were the eight quadrupoles of the high-luminosity insertion at the CERN Intersecting Storage Rings (ISR), strongly focusing the beams around one collision point and more than doubling the luminosity of the collider [11]. The magnets, individually powered, operated in stand-alone liquid helium bath cryostats, fed from a liquefier located some 50 m away via flexible, vapor-screened transfer lines [12]. Such lines, developed together with European industry, have found numerous applications in cryogenic distribution systems and more recently as envelopes of superconducting cables for power transmission. The boil-off from the cryostat baths was used locally to intercept heat on current leads, cryostat neck and thermal shields. Although *a priori* less favorable from a thermodynamic point of view, this choice greatly simplifies the cryogenic distribution, cryostat pipework and control system, thus reducing parasitic heat in-leaks, regaining overall efficiency and improving reliability (all active cryogenic components being located in radiation-free area) [13]. The cryogenic plant made use of turbo-expanders with helium gas bearings, a technology mostly developed in Europe which had by then superseded the piston expansion engines of the old Collins-type refrigerators.

In order to preserve helium purity, the cryogenic plant for the ISR high-luminosity insertion was powered by a dry piston compressor. Shortly after came a revolution in helium refrigeration, with the use of oil-injected screw compressors similar to those employed in higher-temperature refrigeration (figure 5), with the benefits of much lower capital and maintenance expenditure. This was made possible by the development of efficient oil removal from helium, both aerosol and volatile compounds, down to a fraction of ppm level, using three stages of coalescing filters in series with a charcoal adsorber bed [14], a solution implemented on the satellite refrigerators of the Tevatron at Fermilab [15] which soon became the world standard.

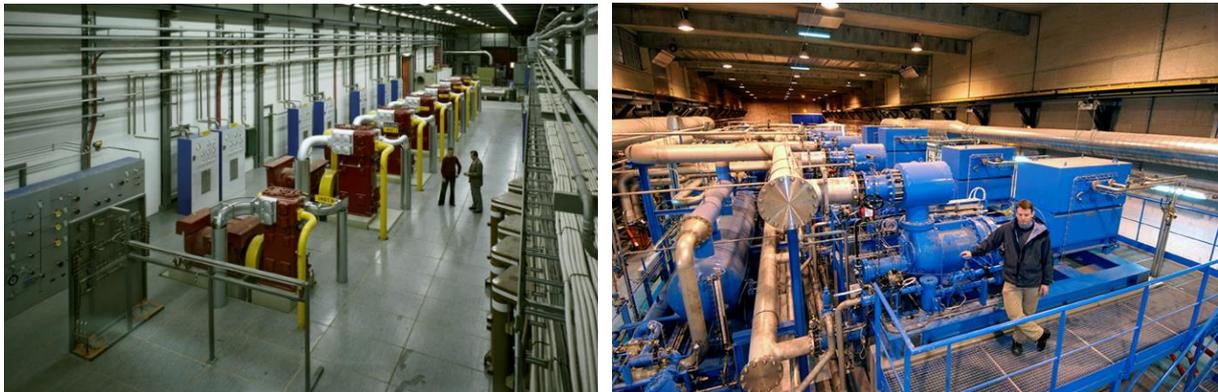

**Figure 5.** Dry piston compressors in the CERN North Area in 1977 (left); oil-injected screw compressors for LEP cryogenics at CERN in 1996 (right)

In those years Brookhaven National Laboratory near New York in the USA had planned to build a 200 to 400 GeV superconducting proton collider called ISABELLE [16]. Magnets were developed, a 3.8 km circumference tunnel was excavated, and a very large helium cryogenic plant was procured from industry with a total capacity of 25 kW at 3.8 K; sub-cooling of saturated helium down to this temperature was achieved by two stages of cold hydrodynamic compressors [17]. The monitoring and control system was computer-based, and the strings of magnets in the tunnel would be cooled by forced-circulation of supercritical helium. The project was eventually cancelled, but the infrastructure, including the cryogenic plant, was later reused for the RHIC collider and has been in operation to this date.

The first fully superconducting particle accelerator was the Energy Doubler/Saver at Fermi National Laboratory near Chicago in the USA, which started operation in 1983 and later became the Tevatron proton-antiproton collider [18]. The 6.3 km circumference tunnel housed 990 main superconducting magnets with a bending field of 4.4 T. The magnets had a room-temperature iron yoke and their coils – non-impregnated to enhance heat transfer to helium – were cooled by forced circulation of supercritical helium, expanded at the end of the magnet sector and continuously re-cooled by the returning two-phase flow of saturated helium in a coaxial annular channel. Liquid helium was produced in a central liquefier and distributed via a cryogenic line circling the ring, to 24 satellite refrigerators operated in economizer mode, each feeding a magnet sector around the ring [19]. A pioneer machine, this successful project established the feasibility of large superconducting magnet systems and their associated cryogenics in accelerators and served as reference for many years. It also trained a whole generation of experts in the construction, operation and maintenance of superconducting magnet and cryogenic systems for particle accelerators, thus preparing the future of the field.

The first superconducting accelerator in Europe was the proton ring of the HERA electron-proton collider at DESY, the German laboratory of particle physics in Hamburg. This machine accelerated beams of protons to 820 GeV, guided around the 6.4 km circumference of the ring by 416 superconducting dipoles and focused by 224 superconducting quadrupoles [20]. The magnets were designed in the laboratory and produced by industrial companies in several European countries. They featured a cold iron yoke and were cooled by forced circulation of supercritical helium at 4.4 K with continuous re-cooling by returning two-phase helium, in a way similar to the Tevatron cooling scheme. The machine was cooled by three large cryogenic refrigerators located in a single hall, each providing 6.8 kW cooling power at 4.4 K, 20.5 g/s liquefaction and 20 kW at 40-80 K for the thermal shields. The refrigerators operated on a modified Claude cycle with 14 heat exchangers and 7 expansion turbines, a configuration optimized for minimizing the temperature pinch on the heat exchange line, achieving a COP of 280 W/W and establishing a new record in thermodynamic efficiency [21]. After the closure of HERA, two of these machines are now re-used, with an additional 2 K stage, for cooling the superconducting high-frequency cavities in the linear accelerator of the European X-FEL project.

In the late 1980s, the USA launched a very large accelerator project, the Superconducting Super Collider (SSC) [22]. After several years of studies and R&D on the superconducting magnets, the construction of the 83 km circumference tunnel for this 40 TeV proton collider started south of Dallas. With almost 10'000 main superconducting magnets in the two proton rings, cost issues were critical and a large effort from several American laboratories aimed at streamlined design, technical optimization and industrialization of the magnets and their cryostats [23]. The project was eventually cancelled in 1993 after several years' construction, but the benefits of the development work proved very valuable for future large projects, including the LHC.

The late 1980s also saw the construction in the USA of the first accelerator making large-scale use of superconducting high-frequency acceleration cavities, CEBAF at the Thomas Jefferson National Laboratory in Newport News [24]. This machine accelerates electron beams by means of two recirculating linear accelerators using niobium superconducting cavities operating at 1.5 GHz, cooled at 2 K in saturated superfluid helium. CEBAF was thus the first high-energy accelerator cooled by superfluid helium, with a single large cryogenic plant nominally producing 4.8 kW at 2 K and 12 kW at 35-50 K for thermal shielding [25]. The powering of the accelerating cavities imposes strong variations on the 2 K heat load, which the cryogenic plant must cope with. Maintaining the saturation pressure corresponding to 2 K on the helium baths of the cavity cryomodules requires four stages of cold hydrodynamic compressors in series and special procedures to handle the temperature and flow transients while maintaining the hydrodynamic machines in their operating envelope, i.e. remaining within the surge, stall and over-speed limits of the wheels. Being the first to produce superfluid-helium refrigeration in the multi-kilowatt range, CEBAF's cryogenic system was extremely useful to the design of that of the LHC. The accelerator started operation in 1995 and is still in use today, after an upgrade to higher energy (and higher cooling capacity of the cryogenic system).

An interesting development occurred in the same period at the Joint Institute for Nuclear Research in Dubna, Russia. A 252 m circumference synchrotron accelerating heavy ions up to 6 GeV/A, the Nuclotron was built using pulsed superconducting magnets. The coils were wound with an internally cooled conductor, so that the magnets rest in the vacuum vessel of the cryostats, without the need for a helium vessel [26]. Magnet cooling circuits are tapped in parallel, between a header supplying liquid helium and a return pipe, and see two-phase flow of strongly varying vapor quality, a configuration prone to generate flow instabilities and possible vapor lock. The system, however, works, thanks to proper balancing of the hydraulic impedances of the parallel circuits and accepting variations in temperature and quality at the outlet of each branch. The latter are not detrimental to the operation of the magnet since they are seen mainly by the iron yoke, cooled in series with the coils. This type of superconducting magnet and cooling scheme is now used in the SIS100 proton and heavy ion synchrotron under construction at GSI Darmstadt, Germany as part of the FAIR complex.

The largest system of superconducting high-frequency cavities built at this date is the acceleration system of the LEP electron-positron collider, built at CERN in the 1980s and upgraded in energy (LEP2) throughout the 1990s [27]. 288 niobium-on-copper cavities, installed deep underground in four long straight sections of the 26.7 km circumference ring, operated at 352 MHz in saturated helium bath cryostats at 4.5 K. They were fed from four large cryogenic helium refrigerators, each with an equivalent capacity of 12 kW at 4.5 K, later upgraded to 18 kW [28]. The upgraded LEP cryogenic plants are now reused for the LHC. They have also set, for several years, a *de facto* standard for the unit capacity of large helium refrigerators, which has seen impressive development since the 1970s following the requirements of the market (figures 6 and 7).

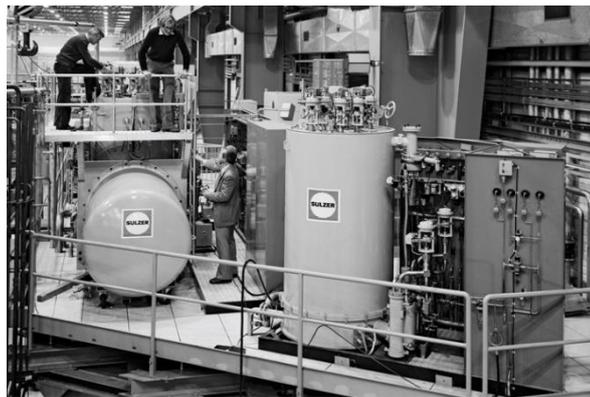

**Figure 6.** 400 W @ 4.5 K refrigerators (SULZER) in CERN North Area in the 1970s

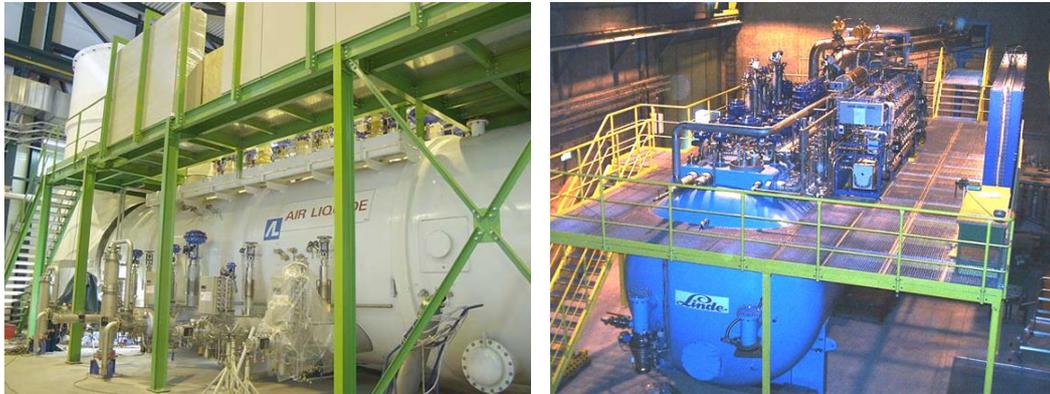

**Figure 7.** 18 kW at 4.5 K refrigerators (AIR LIQUIDE and LINDE) at LEP/LHC in the 1990s

The deep implantation of the LEP tunnel, 100 m below ground on average, sets constraints on the configuration of the cryogenic system. If the large refrigerator cold boxes delivering liquid helium and recovering vapor at 4.5 K had been located integrally at ground level, a reasonable choice dictated by easier maintainability and cost of underground excavation, the hydrostatic head on the column of returning helium vapor of about 20 kPa (200 mbar) would have added 0.2 K on the operating temperature of the cavities. It was then decided to build the cryogenic plants with split cold boxes, a large upper one operating between room temperature and 20 K, and a compact lower one, installed at tunnel level, covering the 20 K to 4.5 K span. Another effect due to tunnel depth also had to be considered, namely the work exerted by gravity on the downward flow of helium, which amounts to some 5 % of the latent heat of vaporization and therefore increases the transfer losses. Oddly enough, this latter effect had been disregarded by one of the cryogenic plant vendors in their first technical offer: application of macroscopic thermodynamics outside usual domains of engineering remains delicate!

The LEP tunnel and technical infrastructure were later reused in the LHC project, the construction of which stretched between 1994 and 2008 at CERN, thus exceeding the time period covered in this Symposium. Prior to construction proper, the LHC however required a decade of study and development work of its critical technologies, namely high-field superconducting magnets and associated helium cryogenics. We now address the development of LHC cryogenics, from first ideas to final implementation.

## 5. Towards the Large Hadron Collider

The first feasibility study of a large hadron collider in the LEP tunnel was presented in a dedicated ECFA workshop in Lausanne in 1984 [29]. It included several options aiming at sounding matter at the level of TeV per elementary constituent, and eventually focused on a proton-proton collider with twin-aperture superconducting bending magnets operating at 10 T, about twice the previous state-of-the-art. This field was deemed achievable either with $Nb_3Sn$ or other A15 superconductors in normal helium at 4.5 K, or the more commonly used Nb-Ti superconductor in superfluid helium at 1.8 K to 2.0 K. The Lausanne workshop had been preceded by several preparatory meetings and included a working group on superconducting magnets and cryogenics, with participation of experts from CEA, CNRS-IN2P3, DESY, INFN, KFK (now KIT), SIN (now PSI) and US laboratories.

The preliminary study of the cryogenic system considered the 4.5 K option, with magnets in bath cryostats integrating distribution piping, definition of operation modes, first estimate of heat loads, eight cryogenic refrigerators distributed around the circumference of the machine each serving two adjacent half-octants, preliminary estimates of cooling capacity, electrical power consumption and cryogen inventory management, and identification of specific safety issues. The alternative option of Nb-Ti or Nb-Ti-Ta magnets in superfluid helium was also mentioned, as deserving further study to permit comparison of both solutions at a later stage.

This was the subject of collaborative work with experts from CEA Grenoble and SIN Villigen (now PSI), resulting in a report [30] presenting a first approach to a pressurized superfluid helium system based on conduction cooling over the length of a half-cell of the accelerator lattice (39.5 m). The need for limiting beam-induced heat loads at 1.8 K was clearly identified, and the report contained first sketches of a normal-helium cooled "beam shield" for intercepting most of them at higher temperature. Two types of solutions were considered for the final stage of refrigeration from 4.5 K down to 1.8 K, namely gas compression and magnetic cycle. The report concluded on the feasibility of superfluid helium cooling, at the cost of increased system complexity and moderate increase in power consumption, thus opening the way for Nb-Ti magnets at high field. The first status report on LHC studies published after the Lausanne workshop [31] is based on this cryogenic scheme.

In the years 1985-1987, discussions between directors G. Brianti (CERN) and J. Horowitz (CEA Institut de Recherche Fondamentale) led the foundation of a collaboration, with two "permanent" working groups on "High-field magnets" (with CEA Saclay) and "LHC low-temperature option" (with CEA Grenoble), the latter to analyze the return of experience from the cryogenics of the superfluid-helium cooled Tore Supra tokamak and to study a forced-convection cooling scheme for the LHC potentially capable of handling higher heat loads. In that period, CERN also launched the "Twin-aperture prototype" (TAP) magnet project, a full-length magnet [32] making use of HERA coils from the then finishing industrial production, to be operated in superfluid helium, for which a 10-m long cryostat was designed and built in industry [33]. A cooperation agreement was signed in October 1988 between directors H. Schopper (CERN) and D. Cribier (CEA-IRF) for the development of technologies for high-field accelerator magnets on one side, and superconducting RF cavities on the other side. Seven execution agreements were conducted under this framework between 1988 and 1994, dealing – as concerns LHC cryogenics – with superfluid helium heat transfer in magnet coils, 1.8 K test stations and cryogenic tests of the TAP, thermo-hydraulics of pressurized and saturated superfluid helium in forced flow, and high-power refrigeration at 1.8 K.

To cope with higher transient and localized heat loads, ill-defined at the time of the first study, a cooling scheme based on forced circulation of superfluid helium was considered [34] and experimentally investigated at CEA Grenoble [35] using a mechanical pump. In parallel, CERN started to investigate a cooling scheme based on two-phase flow of saturated superfluid helium in a distributed heat exchanger pipe traversing the pressurized superfluid helium vessel housing the magnets. A first test set-up at CERN [36] showed the validity of this approach – excellent heat transfer even with limited wetting of the tube wall, thermal cut-off in case of resistive transition thus avoiding quench propagation, avoidance of pumps and of the costly irreversibility and reliability issues associated with them – despite the potential difficulties of two-phase flow in a sloping tunnel [37]. Following this, a cooling loop with full-scale piping was built and operated at CERN under varying regimes, including effects of slope and thermal transients [38], thus opening the way for a full-length prototype magnet string [39]. The appropriate conditions for successful operation of such a cooling scheme – dubbed "helium II bayonet heat exchanger" [40] – were established and the scheme was implemented in the design study of the LHC [41].

Design work also continued on superfluid helium cryostats for housing LHC magnets [42], developing thermal insulation, mechanical construction and leak-tightness techniques suitable for series production, investigating effects of magnet resistive transitions [43] and accidental loss of insulation vacuum [44], and providing experimental input for heat inleak estimates and sizing of the cryogenic system [45, 46]. As the study of the machine matured, dynamic heat loads such as particle losses, synchrotron radiation from the beams, dissipation of image currents in the first wall and resonant acceleration of electrons by the beam potential ("electron cloud") kept evolving, and an interdisciplinary "Heat load working group" was created to track changes and update refrigeration requirements.

Ability for extensive cold testing of full-length prototype magnets triggered the need for a dedicated test bench [47], as the first component of a multi-bench test station designed in modular fashion to be gradually upgraded [48]. Up to twelve test benches could be accommodated, sharing common resources: a 6 kW at 4.5 K helium refrigerator recovered from LEP work, a new room-temperature pumping station providing 120 W at 1.8 K which could be turbocharged by prototype cold compressors to triple the cooling capacity, and a liquid-nitrogen precooling unit enabling the fast cooldown of the large cold mass of the magnets down to 80 K. The magnet test station could thus also serve to test different technologies of cold compressors for the future LHC. Three prototype units were ordered from industry, exploring different technical choices for impeller wheel geometry, motor drive and bearings [49]. It is worth noting that one of these machines, designed and built by a consortium of Czech companies, constituted one of the first supplies from this country after it became CERN member state. All this work formed part of a structured development program for large-capacity refrigeration at 1.8 K [50].

With the successful test of the first full-scale dipole magnet prototype and the approval of the project – conditional to securing extra resources – at end 1994, work entered a new phase. Special in-kind contributions were negotiated with CERN non-member and host states. As concerns cryogenics, this included a collaboration protocol with the French CEA and CNRS, respectively for 1.8 K refrigeration and cryogenic instrumentation, as well as the development and supply of cold compressors from company IHI in Japan, and of the superconducting quadrupoles for the high-luminosity insertions from US laboratories.

As in previous superconducting accelerators, the magnet cryostats originally integrated all cryogenic distribution pipework inside their vacuum enclosure and thermal insulation. The cryogenic layout of the machine was then based on eight cryogenic plants located at the access points of the underground tunnel, each serving two adjacent half octants. Following detailed implantation studies, it was decided to lump the cryogenic plants in five (actually, four plus one) technical sites, thus doubling the flow distribution lengths in the machine tunnel to the 3.3 km of a complete octant. The sizing of the distribution pipework had to be significantly increased, rendering their integration into the magnet cryostat problematic, given the maximum possible transverse occupancy in the tunnel. After a detailed study, the configuration was changed to a separate cryogenic line running next to the magnets, with valve boxes and "jumper" connections every half-cell of the lattice [51]: besides saving valuable transverse space in a crowded tunnel, such a scheme brought the important benefit of decoupling – in technical execution, time schedule and industrial contracting – the installation and reception of the cryogenic line from the magnets. A second prototype magnet string was assembled and tested in this configuration [52], permitting validation of heat loads and thus of refrigeration requirements for the project [53], as well as analysis of magnet quench transients [54]. The cooling scheme was further streamlined [55] by removing one header in the line and halving the number of "jumper" connections to one every cell (106.9 m). In this scheme, the length of the capillary tubes cooling the beam screens was doubled, with the risk of developing instabilities in the heated flow of supercritical helium [56]; a model was built and tested to investigate these effects and establish operating conditions and control strategies for stable flow [57]. Given the novelty, size and complexity of the LHC cryogenic system, as well as its specific conditions of implantation and operation (deep tunnel, limited access, high stored energy, large helium inventory, magnet resistive transitions), a risk analysis process was launched [58] and sustained throughout construction, installation and commissioning.

The configuration and sizing of the LHC cryogenic system were essentially finalized by the end of the millennium [59], allowing launching procurement from industry. Four new cryogenic plants were specified to provide a mix of liquefaction and refrigeration duties at 50-75 K and 4.5-20 K, amounting to total equivalent entropic capacity of 18 kW at 4.5 K [60]. To achieve high thermodynamic efficiency, the adjudication formula was based on the sum of capital price and quoted operating cost over ten years, including electricity and externalities. To close the loop, the effective electrical power consumption would be measured upon reception and compared to the quoted value: a shared incentive bonus/malus would then be applied on the final payment. This approach resulted in excellent efficiency reaching 28 % of the Carnot cycle and settled the old dilemma between capital expenditure and operating cost: although more complex, a more efficient plant can be built smaller – and therefore cheaper – for the same output. In parallel with the procurement of these new installations, the four existing cryogenic plants recovered from LEP were modified and upgraded to the LHC requirements. Procurement was also launched for eight 1.8 K stages producing each 2.4 kW of refrigeration power. The low-pressure helium vapor is compressed by a train of cold hydrodynamic compressors up to a fraction of atmospheric pressure, followed by room-temperature screw compressors operating at sub-atmospheric suction pressure. This arrangement combines good efficiency, limited capital expenditure and compliance with the strongly variable demand resulting from the dynamic heat loads [61]. After functional design and prototyping, the multi-tube cryogenic distribution line running in the tunnel was specified to industry [62]: with a total length of 25.6 km, tight heat inleak requirements (less than 0.2 W/m average on the lowest temperature level) and installation requiring close to 20'000 *in situ* welds, this subsystem represented a formidable technical and financial challenge.

Last of the major cryogenic subsystems, the helium inventory storage and management were defined and their components procured. The LHC contains 135 tons of helium, of which about 60 % are in the magnets when the machine is in operation, the rest being shared between the distribution pipework, the cryogenic plants and the minimum reserve in the buffer storage vessels [63]. Helium is procured from the market and delivered to CERN in standard liquid transport containers. Upon warm-up of the machine, the helium must be stored and its purity preserved. Long-term storage is done at room temperature in 250 m$^3$ gas vessels at 2 GPa (20 bar), which can only accept about half of the inventory. Part of the helium can also be stored, for limited amounts of time, in 120'000-liter vacuum-insulated liquid tanks at atmospheric pressure. The rest is re-injected in the market via "virtual storage" contracts with the gas vendors, a strategy which allows receiving the amounts needed for operation in due time, while limiting the capital expenditure.

## 6. In conclusion

This rapid overview illustrates the impressive development of cryogenics in particle accelerators following that of superconductivity. It shows how a rather exotic laboratory technique has developed into a full-fledged industrial discipline, meeting performance and reliability objectives. Credits for this evolution are due to the pioneering work of the research centers initiating ever ambitious projects (the "science pull"), but also to the specialized industry which has responded to the market demands through sustained technical progress (the "technology push").

**Acknowledgements**

Among them, he would like to particularly acknowledge the work of the CERN Magnet and Cryogenics groups, which he shared for many years.